\documentclass[a4paper]{jpconf}
\usepackage{graphicx}
\usepackage{physics}
\def\be{\begin{equation}}
\def\ee{\end{equation}}
\def\bea{\begin{eqnarray}}
\def\eea{\end{eqnarray}}

\def\f{\frac}

\def\l{\left}
\def\r{\right}

\def\k{\kappa}

\def\d{{\rm d}}
\def\mpl{M_{\rm P}}

\def\7{$\;$}
\def\l{\left}
\def\r{\right}
\def\be{\begin{equation}}
\def\ee{\end{equation}}
\def\bea{\begin{eqnarray}}
\def\eea{\end{eqnarray}}
\def\f{\frac}

\def\k{\kappa}

\def\d{{\rm d}}

\begin{document}
\title{Non-minimal derivative coupling in Palatini cosmology: acceleration in chaotic inflation potential}

\author{Somphoach Saichaemchan, Burin Gumjudpai}

\address{The Institute for Fundamental Study ``The Tah Poe Academia Institute" \\ Naresuan University, Phitsanulok 65000, Thailand}

\ead{buring@nu.ac.th (corresponding)}  \ead{somphoachs58@email.nu.ac.th}




\begin{abstract}
NMDC-Palatini cosmology in slow-roll regime is of our interests.  We present flat FLRW cosmological NMDC-Palatini field equations and acceleration condition. Late time trajectory is approximated. Chaotic inflation potential is considered here as it is viable in a range of positive coupling constant as constrained by CMB results.
Phase portraits show that the NMDC-Palatini gives new saddle-point solutions. In $V \propto \phi^2$ potential, the NMDC-Palatini effect enlarges the acceleration-allowed region from that of GR case.
\end{abstract}
\section{Introduction}
\label{Section:Introduction}
Contemporary cosmology focuses on the either early-universe inflationary \cite{Guth}  or present acceleration expansion hypothesized as effects of dynamical scalar fields such as
inflaton or quintessence \cite{RP1988}, k-essence \cite{Chiba:1999ka} or by effect of gravitational modifications such as braneworlds, $f(R)$, scalar-tensor theories as described in
\cite{AntoShijiLRR2010,Copeland:2006a, maeda,Odintsov:2011,SCVF} and references therein.
On the observation side, the present acceleration and inflation have been convinced by results from SN Ia \cite{Perlmutter:1997zf, Riess:1998cb}, large-scale structure surveys \cite{Tegmark:2004}, cosmic microwave background (CMB) anisotropies
 \cite{arXiv:1001.4538} and X-ray luminosity from galaxy clusters \cite{Rapetti:2005a}.
The prediction suggests that the viable dark energy models for present acceleration should not give too different results from that of the $\Lambda$CDM model while, for the inflation, the model should pass Planck 2015 CMB anisotropies constraint  \cite{Ade:2015lrj}, e.g. giving tensor-to-scalar ratio not exceeding the upper bound of $r < 0.12$ and having spectrum index, $n_s \simeq 0.968 \pm 0.006$.
One way to result in acceleration phase is to have a coupling between the scalar field to gravity sector such as $f(\phi, \phi_{,\mu}, \phi_{,\mu\nu}, \ldots)$
which is motivated in scalar QED, theories with density dependent Newton's constant
\cite{Amendola1993} or with non-minimal derivative coupling (NMDC) terms like  $R \phi_{,\mu}\phi^{,\mu}$
and $R_{\mu\nu} \phi^{,\mu} \phi^{,\nu}$  \cite{Capozziello:1999xt} obtained as lower energy limits of extra-dimensional
theories or Weyl anomaly of $\mathcal{N} = 4$ conformal supergravity \cite{Liu:1998bu, Nojiri:1998dh}.  It is also possible, without  loss of generality, to have coupling term between derivative of
the scalar field and gravity, in form of  $\kappa_1 R \phi_{,\mu}\phi^{,\mu}$  and
 $\kappa_2 R^{\mu\nu} \phi_{,\mu} \phi_{,\nu}$ which as well results in acceleration
\cite{Capozziello:1999uwa, Granda:2010hb}.  Relating the two couplings as $\kappa \equiv   \kappa_2  =  -2 \kappa_1$, combination of the two terms gives the Einstein tensor. Hence this is a theory with $G_{\mu\nu}\phi^{,\mu}\phi^{,\nu}$
\cite{Sushkov:2009, Saridakis:2010mf, Germani:2010gm,  Sushkov:2012,  Tsujikawa:2012mk, Sadjadi:2010bz, Gumjudpai:2015vio, Gumjudpai:2016frh}. The NMC and NMDC are found to be a spacial case of  Horndeski's theory which is generalization of gravitational theory with at most second-order derivatives in the equations of motion, making the Horndeski action the most general scalar-tensor theory \cite{Horn}.  Considering the NMDC theory proposed by \cite{Sushkov:2009},
the metric $g_{\mu \nu}$ is considered as a dynamical field. This approach is the metric formalism. When the metric field and the connection field
both take the role of independent dynamical fields, the approach is of the  Palatini formalism \cite{M, pp1, pp2}. The two formalisms result
in equivalent field equation for the GR case. However, for modified gravity, the field equations alter from each other. The NMDC-Palatini action is expressed as
\bea\label{action NMDC Palatini2}
\tilde{S} = \f{M_{\rm P}^2 c^4}{2}
\int{\d^4x }{\sqrt{-g}} \l[ \tilde{{R}}
-    \l(     \varepsilon   g_{\mu\nu}      +         \kappa\tilde{{G}}_{\mu\nu}     \r)        \phi^{,\mu}\phi^{,\nu}-2V\r]\,,
\eea
of which $\tilde{{R}} = \tilde{{R}}(\Gamma)$,  $\tilde{{G}}_{\mu\nu}(\Gamma)$ and $V=V(\phi)$. The symbol $\varepsilon = \pm 1$ denotes canonical and phantom fields respectively, $M_{\rm P}^2 \equiv (8 \pi G)^{-1}$ and $c=1$.  The matter field term $S_{\rm m}(g_{\mu\nu}, \Psi)$ could be added as well. The Palatini Einstein tensor is
$
\tilde{G}_{\mu\nu}(\Gamma)=\tilde{R}_{\mu\nu}(\Gamma)- ({1}/{2})g_{\mu\nu}\tilde{R}(\Gamma),
$
with details referred to  \cite{Gumjudpai:2016ioy}. In cosmological scenario, the NMDC-Palatini effect, playing a role of dark energy, phantom crossing with oscillating EoS is possible \cite{palaNMDC}. Considering NMDC-Palatini effect, as role of inflaton in the chaotic inflationary model, the tensor-to-scalar ratio and spectral index could pass the Planck 2015 constraint for a range of $\k$ \cite{Gumjudpai:2016ioy}. Note that the NMC Palatini (non-derivative coupling) case was studied by \cite{Odintsov:DE2005}.

\section{NMDC-Palatini cosmology in slow-roll regime}
The NMDC-Palatini field equations in slowly-rolling scalar field regime are  \cite{Gumjudpai:2016ioy}
\be
H^2 \;
   \simeq \;  \f {{\rho}_{\rm tot}}{3 M_{\rm P}^2}
  \l[ 1+\frac{3}{2} \f{\kappa \dot{\phi}^2}{M_{\rm P}^2}(1+w_{\rm eff} )\r]\;\;\;\:{\rm and}\;\;\;  \:
\frac{\ddot{a}}{a}  \simeq  -\f{\rho_{\rm tot}}{6 M_{\rm P}^2}  \l[  1 +\f{7}{2} \f{\kappa \dot{\phi}^2}{{M_{\rm P}^2}}  + 3 w_{\rm eff} \l(1+\f{3}{2} \f{\kappa \dot{\phi}^2}{{M_{\rm P}^2}}  \r) \r],
\ee
with the acceleration condition,
$
w_{\rm eff}   <     - ({1}/{3}) \l( 1 +   {2 \kappa \dot{\phi}^2}/{{M_{\rm P}^2}} \r).
$
Considering only the scalar field as a single species in inflationary epoch, $\rho_{\rm tot} = \rho_{\phi} = {\varepsilon \dot{\phi}^2}/{2} + V$,
$P_{\rm tot}=P_\phi = {\varepsilon \dot{\phi}^2}/{2} - V$ and $w_{\rm eff} =P_{\rm tot}/\rho_{\rm tot}= P_{\phi}/\rho_{\phi}$. Hence
\be
H^2	\simeq \frac{1}{3M_{\rm P}^2}\l[\frac{1}{2}\varepsilon\dot{\phi}^2+V(\phi)  +   \frac{3}{2}\varepsilon   \f{\kappa   \dot{\phi}^4}{M_{\rm P}^2} \r]\,, \label{FRinPhiV1}
\ee
and
\bea
\dot{H} \simeq  \frac{1}{6HM_{\rm P}^2}\l(\varepsilon\dot{\phi}\ddot{\phi}  +  V^\prime\dot{\phi} +  6\varepsilon \f{\kappa\dot{\phi}^3}{M_{\rm P}^2}\ddot{\phi}\r)\;\;\;  {\rm or}\;\;\;
\dot{H} \simeq   -\frac{\varepsilon\dot{\phi}^2}{2M_{\rm P}^2}+\frac{3\kappa\varepsilon\ddot{\phi}\dot{\phi}^3}{4H M_{\rm P}^4}-\frac{3\kappa\varepsilon\dot{\phi}^4}{4M_{\rm P}^4}-\frac{\kappa\dot{\phi}^3V'}{4HM_{\rm P}^4}.\label{HdotExpressionFluid}
\eea
The acceleration is found from the Eqs. (\ref{FRinPhiV1}) and (\ref{HdotExpressionFluid}),
\be
  \f{\ddot{a}}{a}\,  = \, \dot{H} + H^2 \,\simeq\,  -\frac{\varepsilon\dot{\phi}^2}{3M_{\rm P}^2}\l( 1+\frac{3}{4} \f{\kappa\dot{\phi}^2}{M_{\rm P}^2}  \r) + \frac{V}{3M_{\rm P}^2} +\frac{\kappa\dot{\phi}^3}{4HM_{\rm P}^4 }\l(3\varepsilon\ddot{\phi}-V'\r)\,.     \label{eq_acd}
  \ee
The Klein-Gordon equation for slowly-rolling NMDC-Palatini scalar field was derived in  \cite{Gumjudpai:2016ioy}. During inflationary epoch,  $|{\ddot{H}}/{H}|\ll|\dot{H}|\ll|H^2|, \:$
 $|4\dot{H}\kappa| \ll 1,\;$  $ |{9\kappa} \dot{H}/2| \ll 1,\; $ and
$|6\k \dot{\phi}^2/ (5{M_{\rm P}^2})| \ll 1$ so that the equation of motion is hence,
\bea
\ddot{\phi}\simeq\frac{-V^\prime-3H\dot{\phi} \varepsilon }{\varepsilon   -  {(15/2)\kappa}H^2  }\,. \label{phidotdot}
\eea
\subsection{Acceleration condition}
Using (\ref{phidotdot}) in  (\ref{eq_acd}),
\be
 \frac{\ddot{a}}{a}   \simeq  \frac{\l(-{\varepsilon\dot{\phi}^2}   + {V} \r)}{3M_{\rm P}^2}
  -  \l[\frac{\sqrt{3}\kappa V' \dot{\phi}^3}{(4 \sqrt{V} M_{\rm P}^3)} \r] \l\{  \frac{[4 \varepsilon - (5/2) \k V/M_{\rm P}^2 ]}{[\varepsilon - (5/2) \k V/M_{\rm P}^2]}  \r\},\label{eq_acd3}
\ee  where GR part is $\l(   -{\varepsilon\dot{\phi}^2}   + {V} \r)/({3M_{\rm P}^2})$.
The $\dot{\phi}^4$ terms in Eq. (\ref{eq_acd}) and in other equations are negligible. Slow-roll Friedmann equation, $H^2 \approx V/(3 M_{\rm P}^2)$ is used here.
For  $|\k| \ll |M_{\rm P}^2/V| $, binomial approximation is valid. Considering chaotic inflation potential $V = V_0 \phi^n$, this range is
 $|\k| \ll |M_{\rm P}^2 \phi^{-n}/V_0 | $. This agrees with the result in  \cite{Gumjudpai:2016ioy}.
For $\k > 0$, the slowly-rolling scalar field under $V = V_0 \phi^n$ is able to avoid super-Planckian regime. Applying binomial approximation and realizing that the term $\k^2 V^2/M_{\rm P}^4 \ll 1$, the acceleration condition reads,
\bea
 \f{\ddot{a}}{a}\,   \simeq    \f{1}{3M_{\rm P}^2}  \l(   -{\varepsilon\dot{\phi}^2}   + {V} \r)
  -  \frac{\sqrt{3}\kappa V' \dot{\phi}^3}{ \sqrt{V} M_{\rm P}^3}\l( 1 +  \f{15 \varepsilon}{8}\f{\k V }{M_{\rm P}^2}      \r)\, >\, 0.\label{eq_acd4}
\eea
This corresponds to
\be
\varepsilon\dot{\phi}^2 \,<\,  V \l\{1-3\sqrt{3} \l[\frac{\kappa \dot{\phi}^3}{({\sqrt{V}   {M_{\rm P}^2} })}\r]  \sqrt{ 2\epsilon_{\rm v, gr}}\l[1+ \frac{15\varepsilon \kappa V}{({ 8 M_{\rm P}^2})}\r] \r\}, \label{Ac7}
\ee
where $\epsilon_{\rm v, gr} \equiv (M_{\rm P}^2 / 2) (V'/V)^2 $ is a slow-roll parameter.  The Eq. (\ref{phidotdot}) is approximated to  $\dot{\phi} \approx -V'/(3 H \varepsilon)$ and $H \approx \sqrt{V}/(\sqrt{3} M_{\rm P})$ such that $\dot{\phi} \approx -M_{\rm P} V' /(\sqrt{3} \varepsilon \sqrt{V})$, therefore
\be
\varepsilon\dot{\phi}^2 \,<\,  V \l[1 +  \frac{\kappa V' }{\varepsilon  {M_{\rm P}} }  \sqrt{ 2\epsilon_{\rm v, gr}}^3 \l(1+\frac{15\varepsilon}{8}   \f{\kappa V}{ M_{\rm P}^2}\r)\r]. \label{Ac8}
\ee
Positive $\k$ makes it easier to satisfy the acceleration condition than that of GR case (if $V'>0$). Considering chaotic inflation potential, $V = V_0 \phi^2$ with
$V_0 =  \lambda (M_{\rm P}^2)  = (1/2) m^2$, the acceleration condition is
\be
\varepsilon\dot{\phi}^2 \,<\,  V_0 \phi^2  \l[1 +  \frac{16 \kappa V_0 M_{\rm P}^2 }{(\varepsilon  \phi^2) }  + 30 \k^2 V_0^2     \r].
\ee
In the low field (sub-Planckian) region, the second term (with $\k > 0$) could be large. The third term enhances more effect, albeit small amount. It is worth mention the acceleration condition of the NMDC cosmology in metric formalism  \cite{Sushkov:2009}, that is
\be
\dot{\phi}^2 \,<\,  V_0 \phi^2  \frac{\qty[ 1 -  \qty(4 \kappa V_0/[M_{\rm P}(\varepsilon-\kappa V_0 \phi^2/M_{\rm P})]) ]} {\qty[\varepsilon -{3\kappa V_0 \phi^2}/{M_{\rm P}}] }
\ee
(see \ref{AppendixA1}). These regions are shown  in Fig. \ref{fig1}.

\subsection{CMB constraint on $\k$ in chaotic inflation}
 As in \cite{Gumjudpai:2016ioy}, with $\dot{\phi} \approx  - M_{\rm P} V' /(\sqrt{3} \varepsilon \sqrt{V})$, the range of the coupling $\k$ satisfying
 CMB anisotropies constraint: $n_s \simeq 0.968 \pm 0.006$ (Planck 2015 \cite{Ade:2015lrj}), for $V = V_0 \phi^2$ and $\varepsilon = +1$ case, is re-expressed as,
 $
  {0.213}/{m^2} \;  > \; \k  \; > \;    {0.081}/{m^2}\,.
 $
 Coupling is positive and small. Chaotic inflation is viable in this range.

\subsection{Late time trajectory}
Let $\ddot{\phi} \approx 0$ and $\k \dot{\phi}^2/ M_{\rm P}^2 \ll 1$ at late time,
\be
\dot{\phi}\,  \simeq \,    - \frac{V'}{[{3 H(\varepsilon - 4 \dot{H} \k)}]}.
\ee
From Eqs. (\ref{FRinPhiV1}) and (\ref{HdotExpressionFluid}) we approximate that
\be
\dot{H} \simeq - \frac{\varepsilon \dot{\phi}^2}{(2 M_{\rm P}^2)} \qand H \approx \sqrt{\frac{V}{(3M_{\rm P}^2)}} \qty( 1 + \frac{\varepsilon \dot{\phi}^2}{4 V} )
\ee  hence,
\be
\dot{\phi}  \simeq   \frac{- V' M_{\rm P}}{\l\{{\varepsilon\sqrt{3 V} \l[   1  +  \qty(2 \k \dot{\phi}^2/M_{\rm P}^2 )   +  \varepsilon \dot{\phi}^2/(4 V) \r] }\r\}}\,.
\ee
For
 $V = V_0 \phi^2$ potential, the late time trajectory is described by
 \be
 \dot{\phi}  \simeq  \frac{-  2 \sqrt{V_0} M_{\rm P}}{\l\{\varepsilon  \sqrt{3}  \qty[ 1 +  \qty(2\k \dot{\phi}^2/M_{\rm P}^2) + \varepsilon \dot{\phi}^2/(4 V_0 \phi^2)  ]\r\}}\,,
 \ee
  as in Fig. \ref{fig1}.
\section{Autonomous system}
\begin{figure}[h]
\centering
\includegraphics[width=34pc]{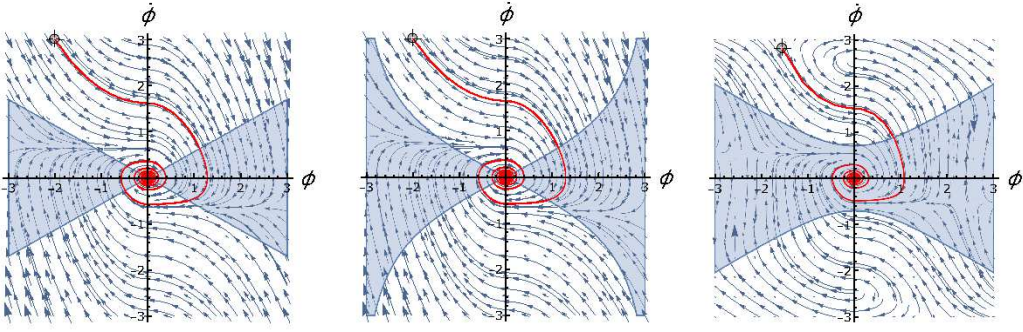}\vspace{1pc}
\caption{\label{fig1} Phase portrait for an NMDC scalar field under $V=V_0 \phi^2$ for $M_{\rm P} = 1.0$ and $m = 0.8$: (a) GR  (b) NMDC-metric formalism ($\k = 0.1$) and (c) NMDC-Palatini formalism
($\k = 0.3$)}
\end{figure}

A three-dimensional closed autonomous system from Eqs. (\ref{HdotExpressionFluid}) and
(\ref{phidotdot}) is formulated:
\be
\dot{\phi}  =    \psi\,,   \;\;\;\;\;\;\;   \dot{\psi}  \simeq    \f{- V'(\phi) - 3 H \psi \varepsilon }{\varepsilon - (15/2) \k H^2}\,,    \;\;\;\;\;\;\;   \dot{H}  =    \f{1}{6 H M_{\rm P}^2} \l(  \varepsilon \psi \dot{\psi}   +  V'(\phi)  \psi  +  6 \varepsilon \f{\k \psi^3 \dot{\psi}}{M_{\rm P}^2}  \r).
\ee
We have phase portrait and acceleration condition in Fig. \ref{fig1} with shadow-labeled region where acceleration is allowed. NMDC-Palatini effect gives new saddle solutions and enlarge acceleration region at small field\footnote{The $\dot{H}$ equation is slow-roll approximated as $\dot{H} \approx V'(\phi)\psi/6H\mpl^2$ in doing the NMDC-Palatini numerical phase plot.}. The GR and NMDC-metric formalism cases are also compared (see \ref{Appendix2}).

\section{Conclusion}
 Field equations, acceleration condition and late-time trajectory of slow-roll scalar field NMDC-Palatini inflation are derived. For a viable chaotic inflation, we give coupling range. GR, NMDC-metric and NMDC-Palatini cases phase portraits are compared. NMDC-Palatini effect gives new saddle and repeller solutions and larger acceleration region at small field.  Stability analysis is awaiting for future works.

\section*{Acknowledgements} Naresuan University Research Grant-R2557C120.

\appendix
\setcounter{section}{0}
\renewcommand{\theequation}{A.\arabic{equation}}
\renewcommand{\thesection}{Appendix \Alph{section}}
\section{Acceleration condition of the NMDC-metric formalism gravity}
\label{AppendixA1}
As of \cite{Gumjudpai:2015vio}, the field equation of the NMDC-metric formalism model with slow-roll approximation is
\begin{equation}
\ddot{\phi}\qty(\varepsilon - 3\kappa H^2) + 3H\dot{\phi}\qty(\varepsilon-3\kappa H^2) + V' = 0,\label{KGEquation1}
\end{equation}
and the NMDC-metric formalism Friedmann equation is
\begin{equation}
	H^2 \simeq \frac{8\pi G}{3}\qty[\frac{\dot{\phi}^2}{2}\qty(\varepsilon-9\kappa H^2)+V ]\,, \label{FMEquation1}
\end{equation}
giving
\begin{equation}
\dot{H}	= \frac{1}{6\mpl^2}\qty[\frac{\varepsilon\dot{\phi}\ddot{\phi}}{H}
			-9\kappa\dot{\phi}\ddot{\phi}H
			-9\kappa\dot{\phi}^2\dot{H}
			+\frac{V'\dot{\phi}}{H}
			].\label{FMEquation1Dot}
\end{equation}

As $H \approx {\sqrt{V}}/{(\sqrt{3}\mpl)}$, $\abs{\kappa}\ll\abs{{\mpl^2}/{V}}$ and chaotic inflation potential, $V(\phi) = V_0 \phi^2$, hence
\begin{equation}
\frac{\ddot{a}}{a}\:\equiv \: \dot{H}+H^2 \, \simeq\, -\frac{1}{3\mpl^2}\qty[\dot{\phi}^2\qty(\varepsilon -\frac{3\kappa V}{\mpl^2})+V]
+\frac{\dot{\phi}V'\sqrt{V}\kappa}{\sqrt{3}\mpl^3},\label{AccelerationEquation}
\end{equation}
assuming $\ddot{\phi}\approx0$ at late time, we can derive late time trajectory as
\begin{equation}
	\dot{\phi} \simeq -\frac{V'}{3H\qty(\varepsilon-3\kappa H^2)}.\label{verylatetimeMetric}
\end{equation}
Using Eq. \eqref{verylatetimeMetric} in Eq. \eqref{AccelerationEquation}, the acceleration condition is
\begin{equation}
\dot{\phi}^2< \left.V\qty[1-\qty(\frac{4\kappa V_0}{\mpl\qty(\varepsilon-\frac{\kappa V_0\phi^2}{\mpl^2})})]\right/\qty[\varepsilon-\frac{3\kappa V_0\phi^2}{\mpl^2}].
\end{equation}
\renewcommand{\theequation}{B.\arabic{equation}}
\section{Autonomous system of the NMDC-metric formalism gravity}
\label{Appendix2}
The autonomous system from Eq.\eqref{KGEquation1} and Eq.\eqref{FMEquation1} are formulated:
\be
\dot{\phi} = \psi \qc \dot{\psi}\simeq -\frac{\qty{V'(\phi)+3H\psi\qty(\varepsilon-3\kappa H^2)}}{\qty(\varepsilon-3\kappa H^2)}\qc \dot{H} \simeq \frac{1}{6\mpl^2H}\qty[\varepsilon\psi\dot{\psi}-9\kappa\psi\dot{\psi}H^2-9\kappa\psi^2\dot{H}+V'(\phi)\psi]
\ee
where $\dot{H}\approx{V'(\phi)\psi}/{6\mpl^2H}$.

\end{document}